# Chemical composition control at the substrate interface as the key for FeSe thin film growth

*Yukiko Obata[1*], Michiko Sato[2], Yuji Kondo[2], Yuta Yamaguchi[3], Igor Karateev[4], Alexander Vasiliev[4,5,6*], Silvia Haindl[1*]*

[1]Tokyo Tech World Research Hub Initiative (WRHI), Institute of Innovative Research, Tokyo Institute of Technology, 4259 Nagatsuta-cho, Midori-ku, Yokohama, Kanagawa 226-8503, Japan

[2]Materials Research Center for Element Strategy, Tokyo Institute of Technology, 4259 Nagatsuta-cho, Midori-ku, Yokohama, Kanagawa 226-8503, Japan

[3]Laboratory for Materials and Structures, Institute of Innovative Research, Tokyo Institute of Technology, 4259 Nagatsuta-cho, Midori-ku, Yokohama, Kanagawa 226-8503, Japan

[4]National Research Centre "Kurchatov Institute," pl. Akademika Kurchatova 1, Moscow, 123182, Russian Federation

[5]Shubnikov Institute of Crystallography of FSRC "Crystallography and Photonics" Russian Academy of Sciences, Leninsky pr. 59, Moscow, 119333, Russian Federation

[6]Moscow Institute of Physics and Technology, National Research University, Dolgoprudny, Moscow region 141701, Russian Federation

**Corresponding authors:**

*Yukiko Obata: obata.y.ab@m.titech.ac.jp

*Alexander Vasiliev: a.vasiliev56@gmail.com

*Silvia Haindl: haindl.s.aa@m.titech.ac.jp

## ABSTRACT

The strong fascination exerted by the binary compound of FeSe demands reliable engineering protocols and more effective approaches towards inducing superconductivity in FeSe thin films. Our study addresses the peculiarities in pulsed laser deposition which determine FeSe thin film growth and focuses on the film/substrate interface, the tendency for domain matching epitaxial growth but also the disadvantage of chemical heterogeneity. We propose that homogenization of the substrate surface improves the control of stoichiometry, texture, and nanostrain in a way that favors superconductivity even in ultrathin FeSe films. The controlled interface in FeSe/Fe/MgO demonstrates the proof-of-principle.

**INTRODUCTION**

The last decade witnessed increasing interest in the thin film growth of tetragonal FeSe (anti-PbO type structure, space group: *P4/nmm*) by a large number of methods ranging from vapor deposition to electrodeposition.[1] The layered van der Waals compound shows a huge flexibility in its electronic band structure and electronic properties upon stoichiometry, strain and doping. This can be easily seen in the ground state. While the superconducting transition for bulk FeSe is in the range of 8 – 9 K,[2] $T_c$ in FeSe films can reach 15 K on CaF$_2$ substrates[3,4] and even 65 – 75 K in FeSe monolayers on SrTiO$_3$.[5–9] FeSe films show a superconductor-to-insulator transition,[10,11] indications for a Berezinskii-Kosterlitz-Thouless (BKT) transition in the ultrathin limit[12] and a response to electrostatic doping as recently shown in electric double layer transistor (EDLT) devices.[13] The underlying superconducting mechanism and a unifying description are still under debate.

In order to investigate the rich physics that FeSe offers, thin film engineering has to provide protocols for its reproducible growth with defined properties. At present, pulsed laser deposition (PLD) encounters limitations in producing ultrathin, superconducting films. While a complete superconducting transition is found in FeSe films on CaF$_2$ substrates even at a thickness of 20 nm,[4] signatures for superconductivity in FeSe films on MgO are either found for films thicker than 100 nm[14] or in 18 nm thin films after post-annealing.[13] Shortly speaking, the choice of the substrate turned out to be a crucial point in the engineering of FeSe films because it has been argued that it affects structural properties (lattice constants, texture) of the films by a strain effect. However, there is a second, less well studied reason: The substrate also affects the chemical composition and the chemical homogeneity of the FeSe films mainly due to interface layer formation or chemical diffusion. In both cases (strain and chemical composition), the electronic properties, in particular the superconducting transition temperature, are affected. Our study is thus motivated by the elaboration of how both aspects are intertwined and complicate the growth of ultrathin FeSe films that should be grown epitaxially and superconducting.

FeSe thin films could be grown with *c*-axis texture (i.e. (001)-orientation) even on non-crystalline materials such as glass as well[15] because van der Waals epitaxy does not necessarily require a strict lattice registry between film and substrate. However, the control of *in-plane* texture does require a crystalline substrate. Here, we have chosen exemplarily to study in detail the growth of tensile strained thin FeSe films on MgO(001) substrates, which generally show suppressed superconductivity. The initial misfit strain $\varepsilon_c$ ($\varepsilon_c = a_{FeSe}/a_{MgO} - 1$) between FeSe and MgO, where $a_{FeSe}$ = 3.77 Å[16] and $a_{MgO}$ = 4.21 Å denote bulk lattice constants of FeSe and MgO at room temperature, is about 12%,[14] which is too large to be accommodated by conventional lattice-matching epitaxy,[17] as depicted in Fig. 1(a). As a result, the FeSe/MgO heterointerface is practically unsuited for coherently epitaxially strained growth. Nevertheless, it is still feasible for FeSe films to grow epitaxially with a *cube-on-cube* texture on MgO (i.e. [100](001)FeSe//[100](001)MgO) because of domain (matching) epitaxy (DME), where *m* unit-cells FeSe match *n* unit-cells MgO across the interface, as described in Fig. 1(b). DME for FeSe/MgO was recently proposed by Harris *et al.*,[18] however with a prediction of *m* = 8 and *n* = 7. Our study clarifies that *m/n* = 11/10, 10/9 or 9/8, fitting perfectly the DME theory by Narayan and Larson applied to a misfit of $\varepsilon_c$ = 10 ± 2 %.[17] Apart from the ideal case of DME for a clean FeSe/MgO heterointerface, we further demonstrate that the real heterointerface is much more complex and chemically heterogeneous due to Fe diffusion into MgO, as shown in Fig. 1(c). We propose that the modified FeSe/Fe/MgO interface, shown in Fig. 1(d), accounts for stabilization of a *cube-on-cube* epitaxy and can lead to lattice-matching epitaxy with a unit cell structure that is advantageous for a superconducting ground state. The results underline the importance of a chemical control in the growth of FeSe films and can be conceptually applied to other substrate choices as well.

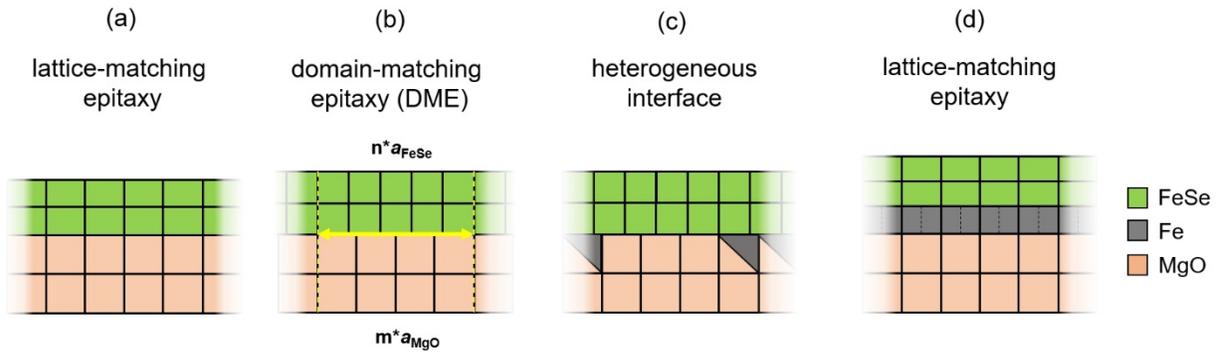

**Figure 1.** Sketch of the FeSe/MgO heterointerface in case of **(a)** ideal lattice-matching epitaxy (cannot be obtained), **(b)** domain matching epitaxy (DME), **(c)** real heterogeneous interface with Fe diffusion, and **(d)** proposed lattice-matching epitaxy by the use of Fe buffer layer in FeSe/Fe/MgO

# EXPERIMENTAL SECTION

**Thin film deposition.** FeSe thin films were grown on single crystalline MgO(001) substrates (10 × 10 × 0.5 mm$^2$, Furuuchi Chemical Co., Japan) by pulsed laser deposition (PLD) using an EV-100/PLD-S growth chamber (Eiko, Japan) under a base pressure smaller than 1 × 10$^{-7}$ Pa. A KrF excimer laser (COMPex Pro 110F, Coherent GmbH, Germany, $\lambda$ = 248 nm) was used for ablating the FeSe target with the laser energy density of 2 Jcm$^{-2}$.

To make the PLD targets, the starting materials of Fe and Se powders were mixed in a stoichiometric ratio and placed in an evacuated silica tube, which was first heated at 400°C for 10 hours and then reacted at 900°C for 24 hours. Subsequently, they were ground, pelletized, and sintered at 900°C for 36 hours. The target composition determined by EPMA was Fe$_{0.99}$Se. The Fe buffers were deposited from a pure Fe commercial target.

The substrates were heated to 600°C in the PLD chamber using an infrared semiconductor laser (LU0915C300-66, Lumics GmbH, $\lambda$ = 915 nm), held at this temperature for 2 hours, and subsequently cooled to deposition temperatures, ranging from 220°C to 500°C. Kept at these temperatures for 30 minutes prior to deposition, FeSe films were fabricated at laser repetition rates of 2 and 10 Hz, respectively. The film thickness was typically 10 – 20 nm. The distance between target and substrate during deposition was approximately 5 cm.

For the deposition of FeSe/Fe/MgO, after the heat treatment of the substrates, they were cooled down to room temperature, where the Fe buffers were deposited at a repetition rate of 10 Hz with a thickness of approximately 10 nm. The Fe-covered MgO was heated to 600°C, held at this temperature for 5 minutes, and subsequently cooled to 250°C. Kept at the temperature for 30 minutes prior to deposition, FeSe films were fabricated at a repetition rate of 10 Hz.

**Characterization.** In order to investigate phase and crystal orientation of the films, we performed standard XRD 2θ/ω in Bragg Brentano geometry and high-resolution scans in parallel beam geometry, using a SmartLab diffractometer (Rigaku) with CuKα radiation. The typical step size Δ2θ of the scans was 0.02°. The *c*-axis lattice parameters were obtained from FeSe(00*l*) reflections with *l* = 1, 2, 3 and 4 by means of a linear extrapolation versus the Nelson-Riley function $\cos(\theta)\cot(\theta)+\cos^2(\theta)/\theta$.[19] The *a*-axis lattice parameters were determined from the peak positions of FeSe(200) reflection in 2θχ/ϕ scans. The film thickness was determined by X-ray reflectivity (XRR) measurements using a SmartLab diffractometer (Rigaku) with CuKα$_1$ monochromated by Ge(220). Film textures and epitaxial relationships were determined from pole-figure measurements using a D8 Discover diffractometer (Bruker, USA) with CuKα radiation.

Surface morphology was characterized by AFM using a MultiMode8 scanning probe microscope (Bruker Nano Inc.) with conventional silicon tips on nitride cantilevers ($f_{res}$ = 130 ± 30 kHz, k = 0.4 Nm$^{-1}$). We evaluated the film surface roughness after flattening the images using the WSxM software.[20] The root-mean-square (rms) roughness was evaluated for 1 × 1 µm$^2$ and 3 × 3 µm$^2$ scans.

Film/substrate interfaces and film cross sections were analyzed by scanning/transmission electron microscopy (S/TEM) and energy dispersive X-ray spectroscopy (EDXS). The specimens for the TEM, STEM and EDXS studies were prepared by using a standard lift-out FIB technique in a Helios Nanolab focus ion beam (FIB) scanning electron microscope (SEM) (Thermo Fisher Scientific, USA). A Pt protective layer with the thickness of 1-2 µm was deposited on the specimen surface by electron beam following Ga$^+$ ion beam deposition. Microstructural analyses were performed in a Titan 80-300 TEM/STEM (Thermo Fisher Scientific, USA) equipped with a spherical aberration corrector (probe corrector) with an accelerating voltage of 300 kV. Such a configuration allows one to obtain images in STEM mode with a resolution of 0.08 nm. The device is equipped with an EDX Si(Li) spectrometer (EDAX, USA), a high-angle annular dark-field (HAADF) electron detector (Fischione, USA) and a Gatan image filter (GIF) (Gatan, USA). In addition, some of the STEM images and EDXS data were obtained in an Osiris (Thermo Fisher Scientific, USA) at an accelerating voltage of 200 kV. This instrument is also equipped with an HAADF detector (Fischione, USA) and a silicon drift detector (SDD), i.e., a Super-X EDX detector (Bruker, USA). Image processing was performed using a digital micrograph (Gatan, USA) and TIA (FEI, USA) software.

Electron probe microanalysis (EPMA) was used for chemical composition analysis of target and films using a JXA-8530F analyzer (JEOL). The elemental analysis was conducted at 50 kV acceleration voltage and an electron current density of $6.3 \times 10^{-3}$ Acm$^{-2}$ (i.e. an electron current of 5 nA on an area of 10 μm in diameter). In addition, Auger electron spectroscopy (AES) depth profiles were acquired on thin films using an ULVAC-Phi 710 Auger electron spectrometer with an integrated scanning electron microscope and an Ar sputtering gun. The analysis was conducted using a focused electron beam with a primary energy of 10 keV and an electron current of 10 nA. The etching rate was ~2 nm·min$^{-1}$ with an Ar$^+$ ion primary energy of 1 keV on a square area of 1 × 1 mm$^2$.

The temperature dependence of the longitudinal resistivity of the films was measured with a physical property measurement system (Quantum Design Inc.) in a range of 2 – 300 K under the magnetic fields of 0 – 9 T. by the four-probe method. Silver paste was employed for electrical contacts. A 90% criterion was used for the determination of $T_c$.

**RESULTS AND DISCUSSION**

**Preparation of FeSe Thin Films by PLD.** A peculiarity of the PLD process is the stoichiometric transfer of material from the target to the substrate. This stoichiometric transfer, however, is violated in the presence of volatile elements. This is the case for FeSe, where the volatility of Se not only limits the substrate temperatures that can be used for thin film growth but also results in a gradual target composition change upon laser irradiation. Starting with a target composition of $Fe_{0.99}Se$ by EPMA we measured a change to $FeSe_{0.38}$ after the growth of ~30 films. In consideration of this strong deterioration in target composition, the target surface needs to be polished more often.

**Structural Characterization.** Fig. 2(a) shows, exemplarily, $2\theta/\omega$ XRD scans for films deposited with the repetition rate of 10 Hz at different substrate temperatures, $T_S = 220 – 500°C$, and illustrates the growth of a *c*-axis oriented, tetragonal FeSe film for a temperature regime between 240 and 500°C (pink shaded region). At lower $T_S$ (220°C), FeSe(110) reflections as well as a secondary $Fe_3Se_4$ phase with (*00l*)-orientation appeared. A similar temperature dependence of the crystalline FeSe phase was also found when films were grown with a repetition rate of 2 Hz (see Supplemental Material Fig. S1). According to EPMA results, the atomic ratio in the deposited FeSe films shows Se deficiency of ~17 at.% (i.e. $Fe_{1.00(2)}Se_{0.83(2)}$). Although the films are Fe-rich, the intensity of the Fe(002) reflection in the $2\theta/\omega$ scans is small. We note that a possible Fe(110) reflection at $2\theta = 44.6°$ is too close to the MgO(002) reflection and may be covered.

Subsequently, the relative *in-plane* orientations were studied for selected films in pole figure measurements based on FeSe(101) and MgO(222) reflections. Compared to the MgO(222) pole figure in Fig. 2(b), the FeSe(101) pole figures in Figs. 2(c) – (g) reveal that the films grew with two different *in-plane* textures: the majority of grains is oriented [100]FeSe//[100]MgO (*cube-on-cube*), a minority shows a [100]FeSe//[110]MgO texture (45º *in-plane* rotation with respect to the majority). A pure epitaxial growth with *cube-on-cube* orientation was only obtained for the film grown at $T_S$ = 500ºC (Fig. 2(g)). We also note that the texture evolution of the FeSe films coincides with changes in crystalline quality. Figs. 2(h) and (i) show the averaged intensities of all pole figure reflections corresponding to the 45º rotated and the *cube-on-cube* textures, $I^{45}$ and $I^{cc}$, respectively, and their dependence on $T_S$. $I^{45}$ is strongest at 260ºC and decreases with increasing $T_S$, whereas $I^{cc}$ increases with increasing $T_S$ reaching most counts at 500ºC. Similarly, the full width half maxima (FWHM) of FeSe(001) reflections from the 2θ/ω-scans in Fig. 2(a) decrease for increasing $T_S$ up to 500ºC (Fig. 2(j)) despite the decreasing peak intensity (Fig. 2(k)).

In contrast to the texture evolution, both *c*- and *a*-axis lattice parameters of the FeSe films show a limited and almost linear temperature dependence (Figs. 2(l) and (m)). The *a*-axis lattice parameters determined by XRD agree well with a valid Poisson effect in both $T_S$-series for 2 and 10 Hz.

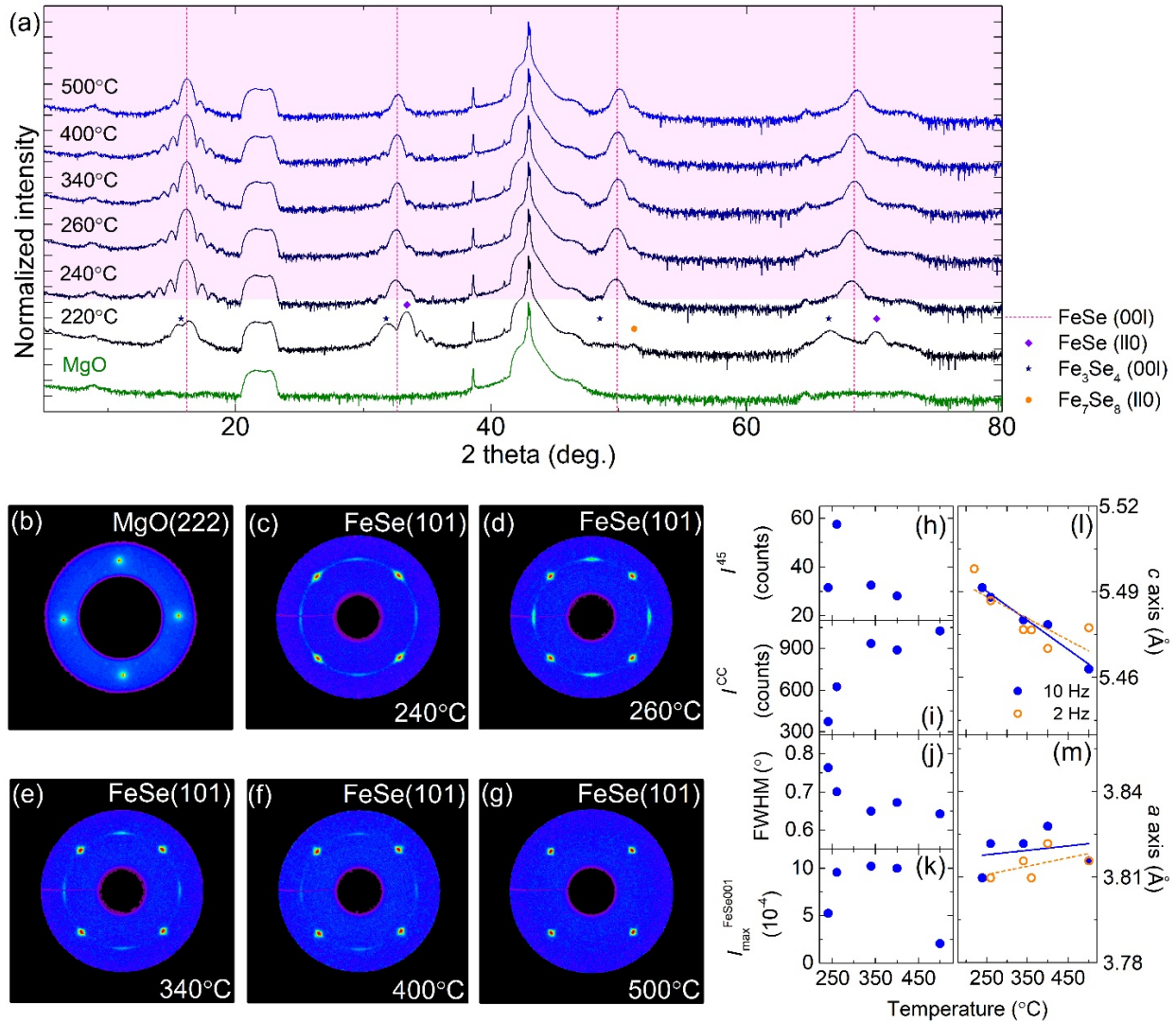

**Figure 2.** XRD results for FeSe films: **(a)** 2θ/ω scans of films and MgO reference (logarithmic scale). Intensities were normalized with respect to the MgO (002) reflection. The pink shaded region shows the *c*-axis oriented FeSe films. **(b)** (222) pole figure of MgO. **(c) – (g)** (101) pole figures of FeSe grown at 240, 260, 340, 400, and 500ºC. The film grown at 500ºC shows a pure *cube-on-cube* texture. **(h) – (k)** Temperature dependence of the average intensities of the peaks corresponding to 45º rotated and *cube-on-cube* textures in FeSe (101) pole figures (**(c) – (g)**) and FWHM and peak intensities of FeSe (001) reflections in XRD 2θ/ω scans. **(l),(m)** Temperature dependence of *c*- and *a*- axis.

**Surface Morphology Analysis.** Additional information is gained by studying the surface morphology of the films by employing AFM. Fig. 3(a) displays the variation of RMS roughness, film thickness, and growth rates with $T_S$. We note an abrupt change in these observables at $T_S$ = 400°C for films deposited at 2 Hz, in contrast to linear evolution for films deposited at 10 Hz. Changes in surface roughness also coincide with those in morphology as demonstrated in AFM images depicted in Figs. 3(b) – (e). Darker areas indicate those locations where the substrates are less covered. They expand more steadily in the films grown at 10 Hz with increasing $T_S$ (Figs. 3(b) – (d)) compared to that deposited at 2 Hz, where the island-like morphology was clearly visible at $T_S$ = 400 and 500°C (Figs. 3(e), S3(e), and S3(f)). These results imply that there is a stronger tendency in the films grown at 2 Hz for a pronounced island growth mode as the substrate temperatures rise. In addition, irrespective of deposition temperatures, the formation of small round precipitates was confirmed on the surface of every film, which grew after exposure to air with time.

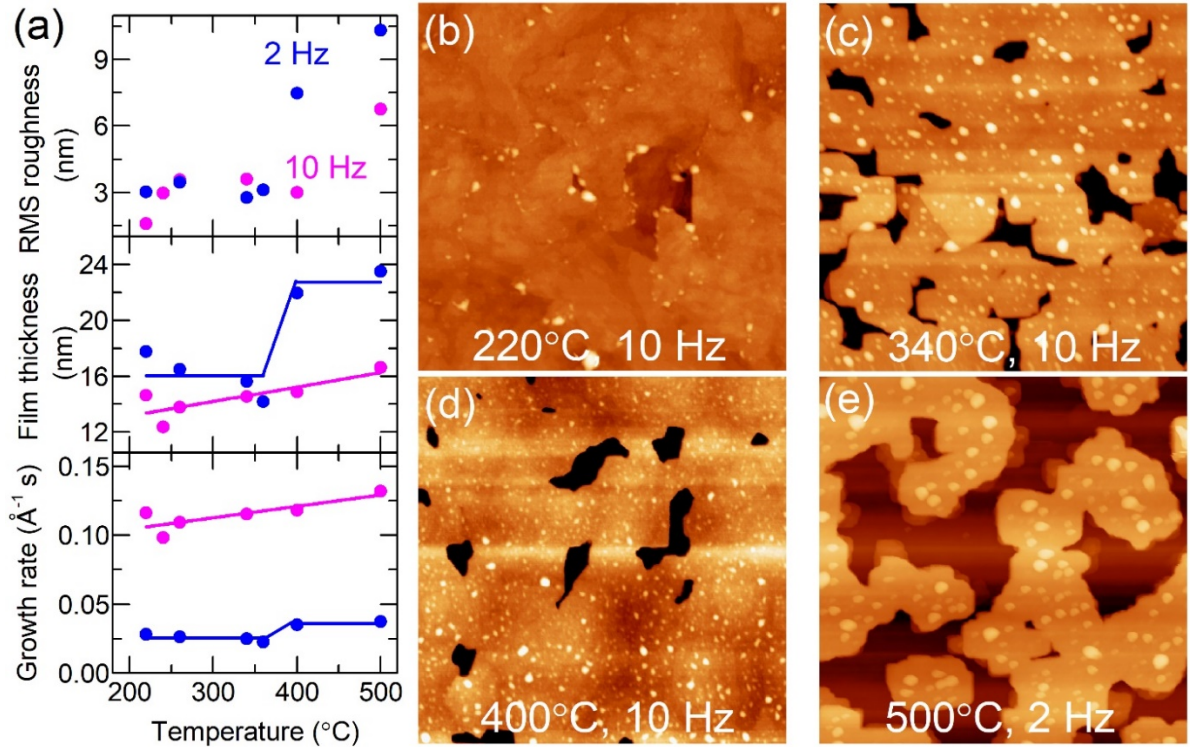

**Figure 3. (a)** RMS roughness, film thickness, and growth rate of FeSe films in dependence of $T_S$. **(b) – (d)** AFM images of the films grown at $T_S$ = 220, 340, and 400ºC with 10 Hz. **(e)** That at $T_S$ = 500ºC with 2 Hz. Fig. S3 shows the AFM images of the full series. All the images were captured on an area of 1 × 1 μm².

**Electrical Transport Properties.** The electrical characterization of the films with thicknesses in the range of 10 – 20 nm agrees with previously reported results of a suppressed superconductivity:[14] all of the films showed a semiconducting-like behavior of their resistivities, and none of the films exhibited superconductivity down to 2 K (see Supplemental Material Fig. S2). Even an additional *in-situ* post annealing of films prepared at 350ºC for 30 minutes did not result in a superconducting transition.

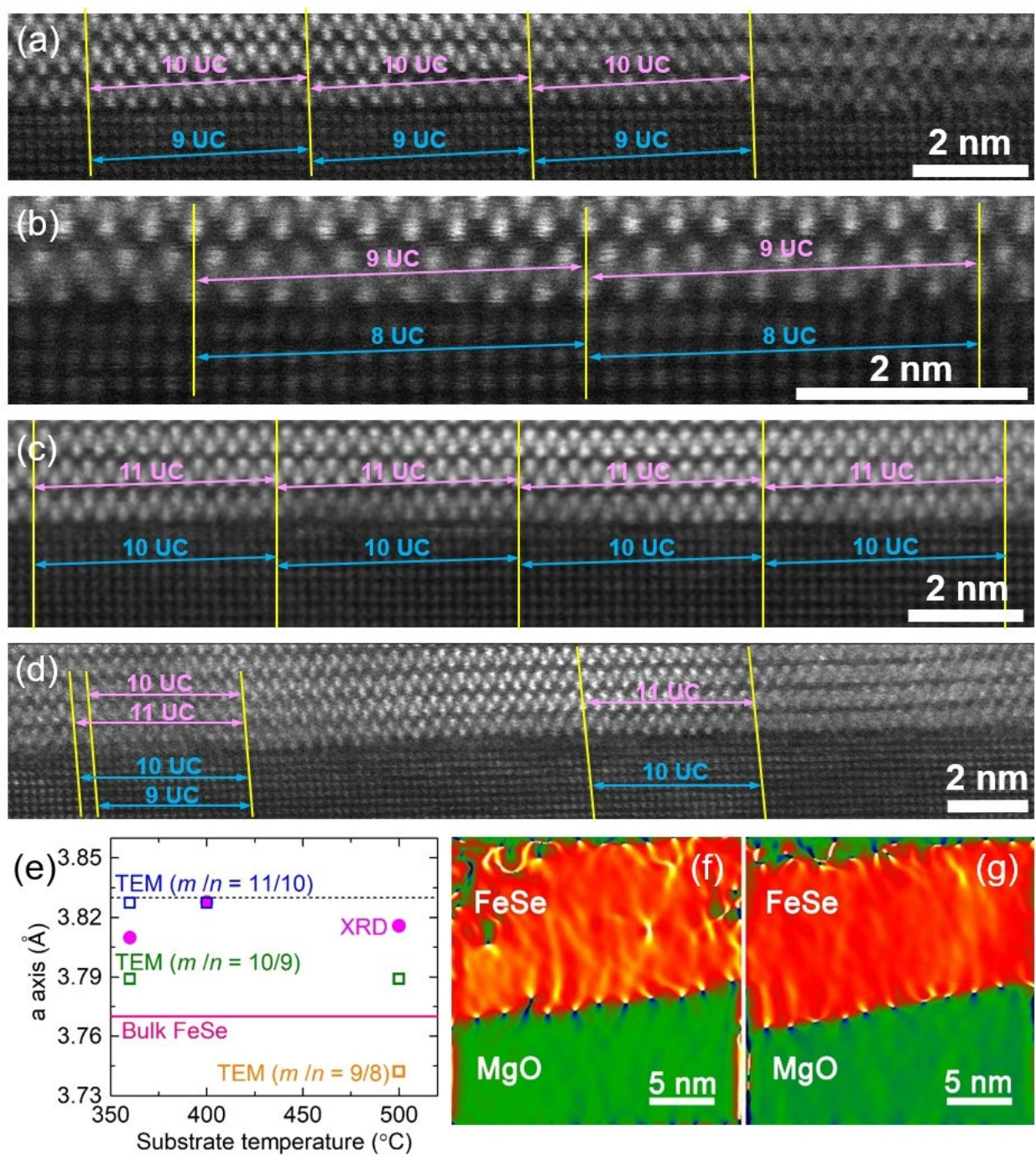

**Figure 4. (a) - (d)** HAADF-STEM images of the FeSe/MgO interface. Yellow lines indicate matching domains of FeSe and MgO unit cells. Note that the FeSe layers in (a) follow a step in the MgO substrate. Film parameters are (a) and (b) 500°C, 10 Hz ($t = 17$ nm), (c) 400°C, 10 Hz ($t = 15$ nm) and (d) 360°C, 2 Hz ($t = 14$ nm). **(e)** Comparison of *a*-axis lattice parameters measured by XRD (pink solid circles) and those obtained from DME (HAADF-STEM) with ratios of $m/n$ = 11/10, 10/9 and 9/8, respectively. The $a$-axis lattice parameter of an FeSe film on MgO reported in Ref.[21] is given by the black dotted line. The pink solid line represents the value of bulk FeSe reported in Ref.[16]. Geometric phase analysis of the FeSe films grown at **(f)** 240°C ($t$ = 13 nm) and **(g)** 400ºC ($t$ = 15 nm). The analysis was performed on an area with cube-on-cube epitaxy.

**STEM Studies of the Interface.** In order to analyze the film/substrate interface in detail, we performed combined STEM and EDS studies on several films. Figs. 4(a) - (d) display the cross-sectional STEM images recorded in high angle annular dark-field mode (HAADF-STEM) of films grown at 500, 400, and 360°C. At the bottom of each image, the cubic lattice of the MgO substrate can be seen, whereas well-ordered FeSe layers are visible on top. The interfaces are a mixture of clean FeSe/MgO and FeSe/Fe/MgO regions. The vertical yellow lines represent the *in-plane* lattice-matching domains. FeSe films have been grown epitaxially on MgO substrates with various domain-matching epitaxial relationships of $m/n$ = 11/10, 10/9 and 9/8, where $m$ unit-cell FeSe match $n$ unit-cell MgO across the interface. Fig. 4(e) compares $a$-axis lattice parameters obtained from DME (HAADF-STEM) with those from XRD results. The variation in domains can be attributed to a heterogeneous interface environment and local variations in lattice strain. Figs. 4(f) and (g) show examples of a geometric phase analysis (GPA) of films grown at 240 and 400°C, where misfit dislocations become easily visible at the film/substrate interface and the color contrast carries information of nanostrain in the film. In both films dislocation cores arrange at the film/substrate interface. The high number of dislocation cores supports the relaxation of the FeSe film lattice. The differences in color contrast suggest that the film grown at $T_S$ = 240°C has a higher strain state than the film grown at 400°C. This is in accordance with the results of the pole figure measurements (Figs. 2(c) – (g)) that show a slightly higher fraction of 45° rotated FeSe domains in the films grown at low $T_S$.

TEM-EDXS is exemplarily shown in Fig. 5(a) for the 14 nm thin FeSe film grown at 360°C. The resulting cross-sectional elemental mapping identifies Fe-enriched areas. In particular, a ~1 nm thin Fe layer formed at the film/substrate interface with an additional influence on the local strain distribution. Furthermore, a ~3 nm thin Fe-rich (or Se deficient layer) was confirmed at the top of the film. In total, an inhomogeneous composition distribution of Fe and Se can be found along the film cross section with Fe-rich zones at the film/substrate interface and close to the film surface. Individual maps for single elements (Fe, Mg, O, Pt, Se) are available in Fig. S5.

**AES Depth Profile Analysis**. AES depth profiling confirms the chemical inhomogeneities along the film cross section of a 31 nm-thin film grown at 350ºC with 10 Hz. The calculated chemical composition with depth (increasing sputtering time) is displayed in Fig. 5(b) and indicates the variation of atomic concentrations of Fe (red), Se (light green), Mg (black), and O (blue). In this example, the gradient of O concentration is broader compared to that shown in the TEM-EDXS elemental mapping (Fig. 5(a)) or in comparable thicker films, and results from the island-like morphology and several observed cracks in the film. Nevertheless, it is clear that a Fe:Se ratio close to 1:1 is only reached in 2/3 of the film cross section, while top regions and the film/substrate interface are Se-deficient. The corresponding first derivatives of AES spectra are shown for selected sputtering times in Fig. 5(c). At a sputtering time of 13 min the Mg KLL transitions appear and indicate the FeSe/MgO interface. Fe LMM intensities remain almost constant across the film, whereas those for Se LMM become weak towards the top surface and the film/substrate interface. We point out that the quality of the films analyzed by TEM-EDXS and AES is comparable.

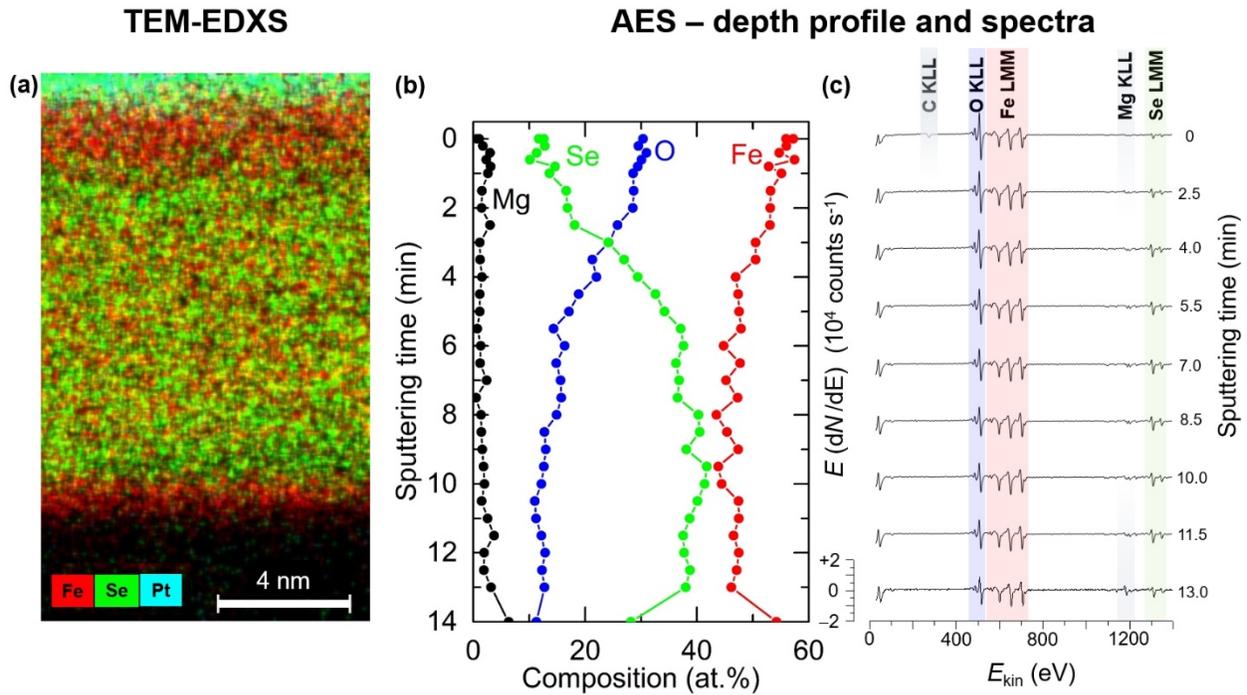

**Figure 5. (a)** TEM-EDXS cross-sectional elemental mapping of the 14 nm thin FeSe film ($T_S$ = 360ºC; rep. rate = 2 Hz). **(b)** AES depth profile (left) and **(c)** derivatives of AES spectra at selected sputtering times (right) of a 31 nm thin FeSe film ($T_S$ = 350ºC; rep. rate = 10 Hz). The gradient of oxygen concentration with respect to depth in AES depth profile is broader than that in the TEM-EDXS elemental mapping, suggesting the larger distribution of FeSe islands which inevitably results in an increase in surface area of the film exposed to air. The films analyzed by TEM-EDXS and AES are qualitatively comparable to each other. Some of the AES depth profiles measured at different positions in the same sample show more Fe-rich concentration along the film cross-section compared to the selected example here.

**Effect of Fe Buffer Layers**. To compensate the chemical inhomogeneity at the interface and minimize the inhomogeneous interface strain, we fabricated FeSe/Fe/MgO films and investigated the effects of Fe buffer layers on texture and electrical transport properties of FeSe/MgO films. Fig. 6(a) shows the resistivity vs. temperature curve of the FeSe/Fe/MgO film ($T_S$ = 250ºC; rep. rate = 10 Hz) at 0 T, exhibiting a complete superconducting transition with $T_{c,on}$ = 3.8 K. From the field dependence of resistivity vs. temperature curves (Fig. 6(b)), we evaluated the field, at which the resistivity reached 90% of its normal-state values at a given temperature, and linearly extrapolated the upper critical field $\mu_0 H_{c2}$ (0) of ~6.8 T, as shown in Fig. 6(c). Fig. 6(d) shows 2θ/ω scans of the FeSe/Fe/MgO film ($T_S$ = 250ºC; rep. rate = 10 Hz) and the FeSe/MgO film grown at a similar deposition condition ($T_S$ = 240ºC; rep. rate = 10 Hz), revealing that the film grown on the Fe-buffered MgO is still *c*-axis oriented. Figs. 6(e) – (g) illustrates the (101) pole figures of FeSe and Fe and (222) pole figure of MgO for the FeSe/Fe/MgO film ($T_S$ = 250ºC; rep. rate = 10 Hz), respectively, confirming the *cube-on-cube* texture of the FeSe film with the epitaxial relationship of (001)[100]FeSe//(001)[110]Fe//(001)[100]MgO.

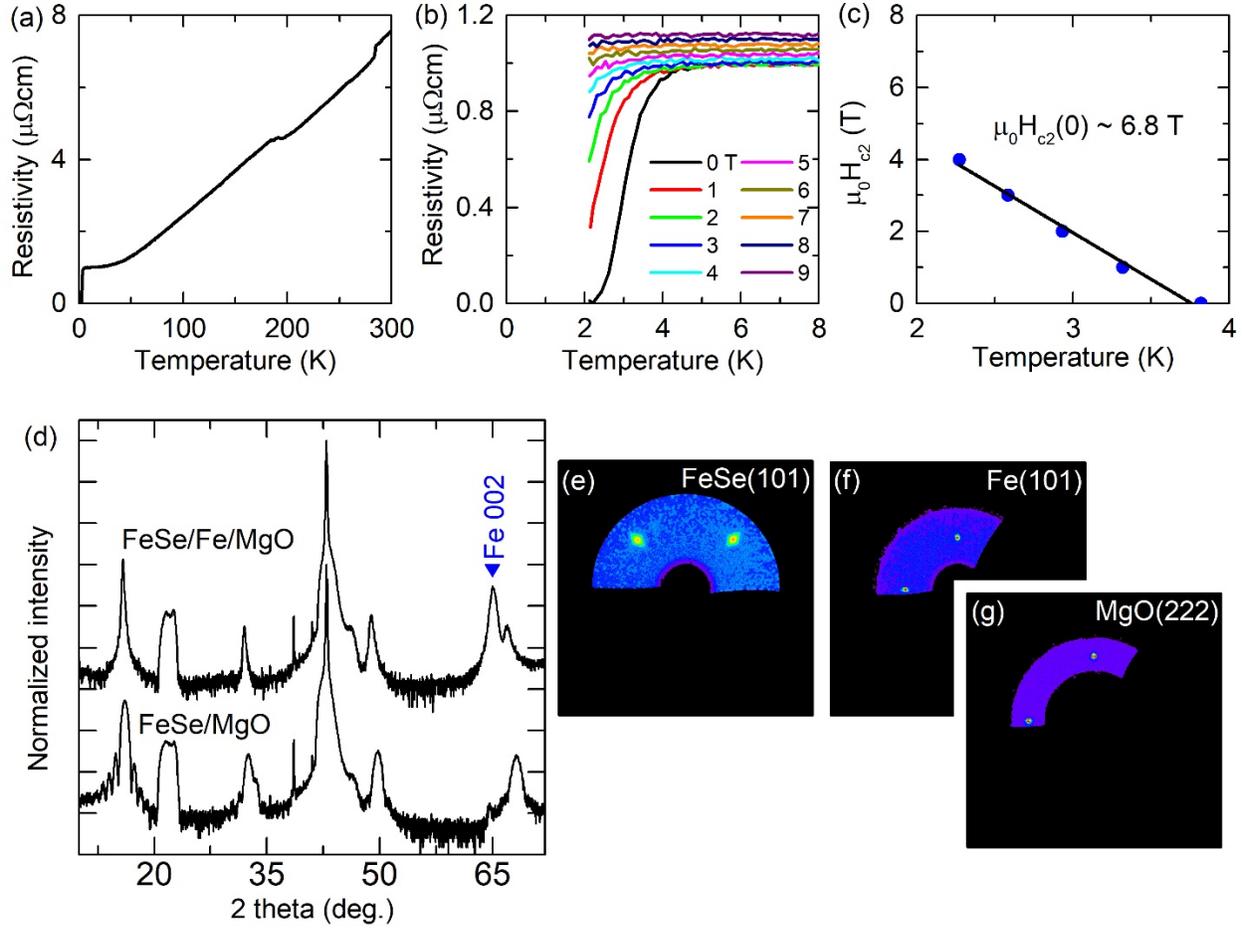

**Figure 6. (a)** Resistivity vs temperature curve for the FeSe/Fe/MgO film ($T_S$ = 250ºC; rep. rate = 10 Hz) at 0 T and **(b)** those at different magnetic fields $\mu_0H//\mathbf{c}$ axis of FeSe. **(c)** Temperature dependence of the upper critical field $\mu_0H_{c2}$ (T) obtained from the fields at which the resistivity reached 90% of its normal-state values at a given temperature. **(d)** 2θ/ω scans of the FeSe/Fe/MgO film ($T_S$ = 250ºC; rep. rate = 10 Hz) and the FeSe/MgO film ($T_S$ = 240ºC; rep. rate = 10 Hz). Blue arrow indicates the Fe(002) reflection. **(e),(f)** (101) pole figures of FeSe and Fe and **(g)** (222) pole figure of MgO for the FeSe/Fe/MgO film ($T_S$ = 250ºC; rep. rate = 10 Hz).

**Discussion.** Several previous studies on the growth of FeSe thin films on MgO[4,14,15,18,22,23] have revealed varying textures of FeSe depending on the growth conditions, predominantly the substrate temperature, $T_S$. Wu et al.[14] first reported a mixture of *cube-on-cube* and 45º *in-plane* rotated textures of FeSe (for $T_S$ = 250 – 500ºC). They attributed the 45º *in-plane* rotation of FeSe unit cells to the large lattice misfit between FeSe and MgO and the thickness dependence of superconductivity to a strain effect. Nie et al.[22] suggested that tensile strain led to the suppression of superconductivity in thinner films of FeSe and, similarly, Wang et al.[23] found a thickness-dependent suppression of superconductivity with respect to various textures.

Harris et al.[18] proposed first that DME applies to the growth of FeSe on MgO. In their study the use of a Se-rich target in PLD (with a laser fluence of 3.4 Jcm$^{-2}$) resulted in the simultaneous epitaxial growth of tetragonal FeSe and hexagonal Fe$_7$Se$_8$ phases (for $T_S$ = 350 – 450ºC). The corresponding epitaxial relationships were found to be [100](001)FeSe//[100](001)MgO, whereas for the (101)-oriented Fe$_7$Se$_8$ phase a mixture of *in-plane* orientations with [010]Fe$_7$Se$_8$//[100]MgO and [010]Fe$_7$Se$_8$//[110]MgO appeared. Higher substrate temperatures (500 – 550ºC) could prevent the growth of Fe$_7$Se$_8$ but supported the growth of (001)- or (101)-oriented tetragonal FeSe with different *in-plane* rotated domains.[18] Important for our study is that for the tetragonal FeSe phase, DME with a ratio of *m/n* = 8/7 was considered to reduce the residual strain $\varepsilon_r$ ($\varepsilon_r = ma_{FeSe}/na_{MgO} - 1$)[17] to 0.35%.[18] The detailed structure of the FeSe/MgO heterointerface remained, however, largely unclear, because high-resolution transmission electron microscopic (HR-TEM) investigations were rare: Although Chen et al.[15] provided evidence of an atomically sharp interface of a 400 nm thin FeSe film grown on MgO at $T_S$ = 320ºC, the origin of the *cube-on-cube* epitaxy despite the large mismatch between FeSe and MgO remained unsolved. A transition layer with a thickness of 1.5 nm within FeSe was mentioned. Zhou et al.[21] has shown a STEM image from the interface of monolayer FeSe films grown on MgO by molecular beam epitaxy (MBE). DME was not investigated, but their image suggests DME with $m = 11$ and $n = 10$. In addition, diffusion of Fe atoms into the top layers of MgO was detected.

The evolution of the crystallographic texture in the thin films shown in Figs. 2(c) – (g) consistently reflects previously noted trends[14,18,24–27] with the difference that the minority fraction of 45º *in-plane* domains is largely reduced. The investigated films have thicknesses below 20 – 30 nm. In comparison to Ref. [18], our study is based on a more stoichiometric target composition. As a result, the Se-rich secondary phase such as $Fe_7Se_8$ does not appear, however, the films are Fe-rich, in particular at the film/substrate interface and towards the film surface. The Fe-excess at the interface leads to a heterogeneous interfacial environment with FeSe/MgO and FeSe/Fe/MgO, having more misfit dislocations and, subsequently, leading to a variation in FeSe domain size. For the apparently clean FeSe/MgO heterointerfaces, we confirmed DME with $m/n = 11/10$, 10/9 or 9/8 by HAADF-STEM. In the case where parts of the substrates are covered by Fe, the epitaxial relation of (001)[110]Fe//(001)[100]MgO as reported in Refs. [28–31], is assisted by the relatively low misfit strain (3.8%). Subsequently, for [100]FeSe//[110]Fe, the initial misfit strain $\varepsilon_c$ between FeSe ($a_{FeSe}$ = 3.77 Å)[16] and Fe ($a_{Fe}$ = 2.87 Å) reduces to 6%. The FeSe/Fe interface could support either conventional lattice matching or DME with increased domains and domain ratios of $m_{FeSe}/n_{Fe} \approx 17/16$.

It is clear that the interfacial structure and its chemical composition have far reaching implications on the electronic properties, primarily on superconductivity. A decrease in *a*-axis lattice parameter accompanied by an increase in *c*-axis lattice parameter is likely to increase the anion height, $h_{Se}$, in the FeSe unit cells and consequently could induce a superconducting transition.

Our results, therefore, accentuate the need for chemically controlled interfaces that could be either a substrate pretreatment (similar to the ones used in the preparation of FeSe monolayers on SrTiO$_3$) or the preparation of an Fe interface layer. Both routes can be regarded as essential for either preventing Se deficiency in FeSe or for minimizing the inhomogeneous interface strain. Both may result in achieving ultrathin superconducting FeSe films by PLD. A first result for FeSe deposition on Fe-buffered MgO demonstrates that a complete superconducting transition with $T_{c,on}$ = 3.8 K is reached for a 19 nm thin epitaxial FeSe film.

**Implications from the Discussion.** *1. Cube-on-cube epitaxy of FeSe on MgO is not realized by conventional lattice matching epitaxy but by DME.* 2. DME occurs on apparently clean FeSe/MgO heterointerfaces due to the large lattice misfit and the experimentally determined domain size matches the DME theory. 3. The real interface between FeSe and MgO is Fe-rich and, therefore, chemically heterogeneous, which impedes the controlled engineering of coherently epitaxially strained FeSe layers whenever the chemical composition of the surface is not controlled. The film surface is covered by condensed Se vapor that forms small round precipitates that chemically react in air, which results in typical artifacts on vapor-deposited films (Figure S6).[32] 4. Misfit dislocations lead to a complex nanostrain pattern in ultrathin films. 5. In addition, Se deficiency (as observed in the films, in particular close to their interfaces) may account for the absence of superconductivity as noted also by Hsu *et al.*[2] 6. The anion height has been in general considered to govern the superconducting transition temperature $T_c$ in Fe-based superconductors.[33] $h_{Se}$ of the FeSe films in the present study was estimated to be ~ 1.5 Å, which is lower compared to 1.73 Å in monolayer FeSe grown on MgO substrates showing a $T_c$ = 18 K.[21] In taking the *c*-axis parameters of FeSe films as a related measure for $h_{Se}$, the decreasing c-axis parameters with increasing $T_S$ are consequently leading to a stronger suppression of superconductivity in the FeSe films, although their *cube-on-cube* texture is improving. 7. Homogenization of the film/substrate interface by an Fe buffer layer which simultaneously supports epitaxial FeSe film growth demonstrates that ultrathin superconducting FeSe films can be deposited by PLD. Further experiments are necessary to explore the thickness limit for superconductivity in this system.

**CONCLUSIONS**

The growth of ultrathin FeSe films using PLD demands the control of the chemical composition at the film/substrate interface. Due to the volatile nature of Se, the PLD process has a tendency in producing Fe-rich films. A detailed microstructural analysis by combining XRD, AFM, AES depth profile analysis, STEM and TEM-EDXS revealed that for apparently clean FeSe/MgO interfaces DME occurs. In addition, we have confirmed that the FeSe/MgO heterointerface is heterogeneous and is Fe-enriched leading to a mixture of FeSe/MgO and FeSe/Fe-O/MgO. A drawback of the Fe diffusion into MgO is that it complicates strain analysis and might be responsible for the variety of nanostrain patterns found by GPA, making the properties of FeSe films less controllable. Furthermore, in improving the film epitaxy by increasing the substrate temperature during deposition is simultaneously driving the structural parameters of FeSe (tetrahedral bonding angle) into a direction that is unfavorable for superconductivity. The above results underline the general importance of a chemical control of the substrate surface in PLD of FeSe films.

## ASSOCIATED CONTENT

**Supporting Information**. The Supporting Information is available free of charge on the ACS Publications website at DOT:. brief description (PDF)

## Author Contributions

Y. O. has grown and characterized the thin films, analyzed data, and wrote the manuscript; Y. Y. prepared the FeSe target for PLD experiments; M. S. performed AES depth profiles and their analysis; Y. K. conducted the EPMA experiments and SEM imaging of the target surface; I. K. performed sample preparation for the TEM/STEM and conducted STEM study; A. V. conducted and analyzed the TEM/STEM/EDXS studies; S. H. conceived and designed the study, interpreted the data, and wrote the manuscript. All authors read and approved the final submitted manuscript.

## ACKNOWLEDGMENT

The work was supported by WRHI Tokyo Tech and partially supported by the Ministry of Education, Culture, Sports, Science and Technology (MEXT) through Element Strategy Initiative to Form Core Research Center. Y. O. thanks Kota Hanzawa, Hidenori Hiramatsu and Hideo Hosono.